# Fundamental Survey on Neuromorphic Based Audio Classification

Amlan Basu [1], Pranav Chaudhari [2], Gaetano Di Caterina [1]

[1] Neuromorphic Sensor Signal Processing Lab, Centre for Image and Signal Processing,
Electrical and Electronic Engineering Department,
University of Strathclyde, Glasgow, U.K.

*Abstract*—Audio classification is paramount in a variety of applications including surveillance, healthcare monitoring, and environmental analysis. Traditional methods frequently depend on intricate signal processing algorithms and manually crafted features, which may fall short in fully capturing the complexities of audio patterns. Neuromorphic computing, inspired by the architecture and functioning of the human brain, presents a promising alternative for audio classification tasks. This survey provides an exhaustive examination of the current state-of-the-art in neuromorphic-based audio classification. It delves into the crucial components of neuromorphic systems, such as Spiking Neural Networks (SNNs), memristors, and neuromorphic hardware platforms, highlighting their advantages in audio classification. Furthermore, the survey explores various methodologies and strategies employed in neuromorphic audio classification, including event-based processing, spike-based learning, and bio-inspired feature extraction. It examines how these approaches address the limitations of traditional audio classification methods, particularly in terms of energy efficiency, real-time processing, and robustness to environmental noise. Additionally, the paper conducts a comparative analysis of different neuromorphic audio classification models and benchmarks, evaluating their performance metrics, computational efficiency, and scalability. By providing a comprehensive guide for researchers, engineers and practitioners, this survey aims to stimulate further innovation and advancements in the evolving field of neuromorphic audio classification.

## I. INTRODUCTION

Sound classification is a critical task in a spectrum of applications, including, but not limited to, surveillance, healthcare monitoring, and environmental analysis. Traditionally, the efficacy of sound classification methodologies has been constrained by the limitations inherent in conventional approaches, characterized by complex algorithms and manual feature engineering. These methods often struggle to encapsulate the intricate nuances of sound patterns, leading to suboptimal performance in real-world scenarios. However, the emergence of neuromorphic computing [1] offers a transformative paradigm shift in this domain. Inspired by the intricate architecture and functionality of the human brain, neuromorphic systems possess the potential to revolutionize sound classification by mimicking the brain's ability to process information in parallel and adapt to dynamic environments. This review delves into the revolutionary domain of neuromorphic-based sound classification and aims to provide a comprehensive overview of the current state-of-the-art in neuromorphic-based sound classification, examining foundational principles, key components, methodologies, and practical applications. By bridging the gap between neuroscience-inspired computing and sound processing, this work endeavors to pave the way for further innovation and advancement in this rapidly evolving field [2].

Neuromorphic computing, inspired by the brain's architecture and functioning, offers a promising alternative by providing inherently parallel and adaptive processing capabilities, which are well-suited for handling such complexities. Secondly, Neuromorphic computing holds the potential to address key limitations of traditional approaches, including high computational costs, limited scalability, and inefficiencies in dealing with dynamic or noisy data. By mimicking the brain's ability to process sensory information in real-time and adapt to changing conditions, Neuromorphic systems offer the prospect of more efficient and robust sound classification algorithms. Moreover, the review of sound classification using Neuromorphic data is timely due to recent advancements in Neuromorphic hardware and algorithms, which have significantly enhanced the feasibility and effectiveness of applying Neuromorphic approaches to real-world problems. By providing a comprehensive overview of the state-of-the-art techniques, methodologies, and applications in this field, such a review can serve as a valuable resource for researchers, engineers, and practitioners seeking to leverage Neuromorphic computing for sound classification tasks. Neuromorphic computing has the potential to enable disruptive innovation in sound classification, offering more efficient, robust, and adaptive systems compared to traditional approaches. This review highlights the transformative potential of Neuromorphic computing in revolutionizing sound processing technologies and outlines avenues for harnessing this potential to address real-world challenges. Overall, the importance of this review lies in its synthesis of cutting-edge research, its timely assessment of advancements in Neuromorphic computing, and its potential to guide future research and innovation in the field of sound classification [3].

The paper is arranged as follows. Firstly the fundamental principles of SNNs, including neuron models and synapses, are presented in Section II. Then, the main properties of biological neural networks are detailed and the results of main works that are inspired by the working mechanisms of biological neural networks are summarized, in Section III. After that, the discussion section tabulates the results obtained and compares it to previous works (Section V). Moreover, this section also includes a discussion of the results and an assessment of the impact of the research. Finally, Section VI concludes the paper, where the limitations and future work are discussed.

## II. FUNDAMENTALS OF NEUROMORPHIC ENGINEERING

### A. Spiking Neural Networks

The concept of artificial neural networks (ANNs) is rooted in biology, as ANNs have evolved to mimic the physical connections between the neurons in the brain, which communicate to perform functions like information processing and decision-making. However, although resembling their biological counterparts, the artificial neurons employed in ANNs do not precisely replicate their behavior. This discrepancy has prompted the emergence of Spiking Neural Networks (SNNs). Unlike biological neural networks, where impulses are not directly transmitted between neurons but instead involve neurotransmitter exchange, SNNs aim to closely mimic this mechanism. Thus, rather than operating with continuously changing values over time like ANNs, SNNs function with discrete events at specific time points, known as spike trains. At any given moment, each neuron within an SNN holds a value akin to the electrical potential of biological neurons. This value can fluctuate based on the neuron's mathematical model; for instance, receiving a spike from an upstream neuron may cause it to increase or decrease. If this value exceeds a set threshold, the neuron will emit a single impulse to connected downstream neurons. Consequently, the neuron's value rapidly decreases below its average, similar to the refractory period experienced by biological neurons, before gradually returning to its average state over time [1].

Let $i$ represent a post-synaptic neuron, where $u_{i,t}$ represents its membrane potential, $o_{i,t}$ stands for its spiking activation, and $\lambda$ denotes the leak factor. The index $j$ refers to the pre-synaptic neuron, and the weights $w_{i,j}$ govern the synapse values connecting these neurons [4]. The iterative update of the neuron activation is computed as follows:

$$o_{i,t} = g\left(\sum_j (w_{i,j} o_{j,t}) + \lambda \cdot u_{i,t-1}\right) \quad (1)$$

The function $g(x)$ in (2) acts as a thresholding mechanism, transforming voltage into spikes.

$$g(x) = \begin{cases} 1, & x \geq U_{th} \\ 0, & x < U_{th} \end{cases} \quad (2)$$

Following the spike, a reset occurs by subtracting $U_{th}$ from the current membrane potential $u_{i,t}$, resulting in $u_{i,t}$ representing the membrane potential after resetting.

### B. Models of SNN

*1) Hodgkin-Huxley model:* The Hodgkin-Huxley model [4] is a fundamental mathematical framework used to understand the generation and propagation of electrical signals in neurons, particularly in the context of action potentials. Proposed by British scientists Alan Hodgkin and Andrew Huxley in 1952, this model describes how the movement of ions across the neuronal membrane leads to changes in the cell's electrical potential, ultimately resulting in the generation of an action potential. In essence, the Hodgkin-Huxley model depicts neurons as complex electrical circuits with specialized channels for different ions such as sodium, potassium, and chloride. These channels open and close in response to changes in the membrane potential, allowing ions to flow across the membrane in a highly controlled manner. This flow of ions creates electrical currents that drive the dynamics of the neuron's membrane potential over time. The model consists of a set of differential equations that describe the dynamics of ion currents and the membrane potential. It incorporates parameters such as membrane capacitance, ion conductances, and reversal potentials, to accurately capture the behavior of real neurons. One of the key insights provided by the Hodgkin-Huxley model is the understanding of the mechanisms underlying the initiation and propagation of action potentials. By simulating the dynamics of ion channels and membrane potential changes, the model can predict how neurons respond to various stimuli and how signals are transmitted along their length.

In the Hodgkin-Huxley (HH) model, the neural membrane is conceptualized as an electrical circuit. The membrane's lipid bilayer is represented as a capacitor, denoted by $C$, with the membrane potential $V$ serving as the voltage across this capacitor. Ion channels on the membrane, such as those for sodium, potassium, and leakage current, are likened to resistors, designated as $R_N$, $R_K$, and $R_l$ respectively. $R_N$ and $R_K$ resistances may exhibit time-dependent variations, whereas $R_l$ remains constant. Each type of ion possesses an equilibrium potential due to concentration disparities across the membrane. When the membrane potential equals the equilibrium potential of a specific ion, the resultant electric current arising from the movement of ions of that type becomes zero. The equilibrium potentials for sodium, potassium, and leakage current are denoted by $E_N$, $E_K$, and $E_l$, respectively. Upon reaching a certain threshold, the membrane potential triggers the neuron to emit an output spike, known as an action potential, which propagates to downstream neurons before the membrane potential swiftly resets. Following an output spike, the neuron undergoes a refractory period during which it is unresponsive to additional input spikes. The HH model can manifest as point neuron models [5] or more intricate multi-compartment neuron models. In multi-compartment neuron models, the membrane's different segments (compartments) may possess distinct photoelectrical properties. Each compartment can thus be represented by a separate set of differential equations incorporating unique parameter values. The HH model demonstrates remarkable fidelity in replicating experimental observations from neurophysiological experiments, contingent upon the accurate setting of parameters. Nonetheless, the HH model's computational demands are substantial, leading to simulations that are typically hundreds of times slower than in real time. It should be noted that biological neural networks are far more complex than even the most detailed multi-compartment HH model, illustrated below (among many others):

- Dendritic Spikes: Dendrites can independently generate spikes and convey them to the soma. The occurrence of a dendritic spike, contingent upon its timing and location, can either elicit or suppress a complete spike at the soma. Given that the average neuron boasts approximately 10,000 synapses, a neuron adorned with myriad synapses exhibits formidable learning prowess akin to that of an entire neural network comprising point neurons.
- Glia Cells: Traditionally regarded as providers of support and protection for neurons, glia cells have often been overlooked by many models in neuromorphic computing. However, recent revelations by neuroscientists indicate that glia cells may hold a pivotal role in information processing and communication. Consequently, they cannot be disregarded in pursuit of accurate modeling.
- Gene Regulation Triggered by Neurotransmitters: Despite possessing identical DNA to most cells in an organism, neurons typically harbor a distinct genome. This distinct genetic makeup enables neurons to dynamically modify their genome, catering to the exigencies of diverse information processing tasks, thereby facilitating self-adaptation.
- Electrical Synapses: In addition to synapses reliant on chemical neurotransmitters, certain synapses engage in communication through electrical signals.

*2) Leaky Integrate and Fire Model:* The leaky integrate-and-fire (LIF) model [4] is a simplified mathematical framework used to describe the behavior of neurons in response to incoming electrical signals. In this model, each neuron is conceptualized as an electrical circuit with a leaky capacitor and a threshold for firing. At its core, the neuron's membrane potential is represented as a capacitor that leaks charge over time, akin to a leaking bucket gradually losing water. When the neuron receives input signals, typically in the form of synaptic currents, these signals are integrated or summed over time by the membrane potential. This integration process reflects the neuron's tendency to accumulate electrical charge in response to incoming signals. Once the membrane potential reaches a certain threshold, the neuron fires an action potential, or spike, signaling the transmission of information to downstream neurons. This firing event is akin to the overflowing of the leaky capacitor, triggered by the accumulated charge exceeding a critical level. Following a spike, the membrane potential is reset to a resting state, reflecting the discharge of accumulated charge and the refractory period during which the neuron is temporarily unresponsive to further input. Despite its simplicity, the LIF model captures essential aspects of neuronal behavior, including the integration of synaptic inputs and the generation of action potentials. It serves as a valuable tool for understanding the dynamics of neural networks and for simulating large-scale neuronal systems in computational neuroscience research. In LIF model, the neuron will accumulate the potential from the input, once its potential reaches the threshold, the neuron will be fired with a spike [6].

*3) Izhikevich Model:* The Izhikevich model [4] represents a simplified yet versatile mathematical framework for describing the dynamics of neuronal activity. It was introduced by Eugene Izhikevich as an alternative to more complex models like the Hodgkin-Huxley (HH) model, aiming to strike a balance between biological realism and computational efficiency. The Izhikevich model conceptualizes neurons as dynamic systems governed by a set of nonlinear differential equations. Unlike the HH model, which involves numerous parameters and equations to capture the intricate behavior of ion channels, the Izhikevich model simplifies neuronal dynamics into two equations: one governing the membrane potential and another controlling the recovery variable. Despite its simplicity, the Izhikevich model can replicate various neuronal behaviors observed in biological systems. For example, it can mimic the generation of action potentials, the refractory period following spiking activity, and various firing patterns such as regular spiking, bursting, and fast-spiking. In terms of resemblance to the HH and LIF models, the Izhikevich model shares some similarities. Like the HH model, it can capture nuanced neuronal behaviors and firing patterns, albeit with a simpler mathematical formulation. At the same time, it retains the computational efficiency of the LIF model, making it suitable for simulating large-scale neuronal networks and exploring complex dynamics in neural circuits. The selection of modeling abstraction levels should align with the intended purpose of the simulation. To create efficient simulation tools for neuroscience research, employing more biologically realistic neuron models, such as the HH model, is advisable despite its computational demands. Conversely, if the aim is to develop information processing algorithms and chips based on SNN, simpler neuron models like the LIF, adaptive exponential integrate-and-fire (AdExIF) [7], or Izhikevich models suffice for the task. In this context, additional complexities introduced by more intricate models, such as the HH model, often do not contribute significantly to improving algorithm performance [8]. Overall, the Izhikevich model offers a valuable compromise between biological accuracy and computational tractability, making it a popular choice for studying neuronal dynamics and network behavior in computational neuroscience research.

## III. Survey of Sound Classification using SNN

Sound classification using Spiking Neural Networks (SNNs) involves a different approach compared to traditional artificial neural networks. SNNs are inspired by the way neurons communicate in the brain, where information is encoded in the timing of spikes, or action potentials, rather than in continuous firing rates. In sound classification, SNNs process audio signals in a manner analogous to how the auditory system operates in the brain. The audio signal is first converted into a format suitable for neural network processing, such as a spike train representation. This conversion can be achieved using techniques like spike encoding, where spikes are generated based on the amplitude and timing of the audio signal. Once the audio signal is encoded into spikes, it is fed into the input layer of the SNN. The SNN then processes these spikes through layers of neurons, with connections between neurons representing synapses. The neurons integrate incoming spikes over time and produce output spikes based on certain firing thresholds. Training an SNN for sound classification typically involves spike-based learning algorithms, such as Spike-Timing-Dependent Plasticity (STDP) [4], which adjust the synaptic weights between neurons based on the timing of pre- and postsynaptic spikes. Through this learning process, the SNN learns to distinguish between different sound classes by adjusting its synaptic weights to maximize classification accuracy. One advantage of using SNNs for sound classification is their potential for low-power and event-driven processing, which can be advantageous for energy-efficient implementations in resource-constrained devices. Additionally, SNNs have been shown to exhibit robustness to noise and variability, making them suitable for real-world audio environments. Overall, sound classification using SNNs represents an exciting avenue for research, offering a biologically-inspired approach to audio processing with potential applications in areas such as speech recognition, environmental sound analysis, and auditory scene understanding.

### A. Overview of Transformer-Based SNN

Martinelli et al. [9] presents a novel approach to implementing VAD using SNNs trained with backpropagation, aiming for low-power neuromorphic systems. The proposed system leverages the inherent parallelism and event-driven processing nature of SNNs, which can lead to significant reductions in power consumption compared to traditional digital implementations. The core of the proposed method lies in the training process of SNNs using backpropagation, a technique widely employed in conventional artificial neural networks (ANNs). Backpropagation enables the optimization of SNN parameters to accurately classify input audio signals as either containing voice activity or being silent. By iteratively adjusting synaptic weights based on the error between predicted and actual outputs, the network learns to discriminate between voice and silence events effectively. One key advantage of using SNNs for VAD is their ability to process input signals in real-time while consuming minimal power. Unlike traditional VAD algorithms that operate on frames of audio data, SNNs can operate asynchronously, reacting to input events as they occur. This event-driven processing enables efficient utilization of computational resources, further reducing power consumption. The authors experimentally validate the proposed approach using benchmark datasets and demonstrate its effectiveness in accurately detecting voice activity with low power consumption. By leveraging the capabilities of neuromorphic hardware platforms, such as spike-based processors or specialized neuromorphic chips, the proposed SNN-based VAD system shows promise for deployment in energy-constrained environments, such as mobile devices or Internet of Things (IoT) devices. Furthermore, the paper discusses potential extensions and optimizations to enhance the performance and efficiency of the proposed system. This includes exploring network compression and quantization techniques to reduce the hardware footprint and investigating alternative training algorithms tailored for spiking neural networks. In conclusion, integrating spiking neural networks trained with backpropagation offers a promising avenue for implementing voice activity detection with low power consumption in neuromorphic hardware. This research contributes to advancing energy-efficient audio processing systems, with potential applications in various domains, including speech recognition, voice-controlled devices, and intelligent sensor networks.

Peterson et al. [10] introduces a novel method that combines Spike-Timing-Dependent Plasticity (STDP) with back-propagated error signals for training SNNs specifically tailored for audio classification tasks. STDP is a biologically inspired learning rule where synaptic strengths are modified based on the relative timing of pre- and post-synaptic spikes. While STDP is well-suited for unsupervised learning in SNNs, it cannot handle supervised tasks such as audio classification, where explicit error signals are needed to guide learning. The proposed approach addresses this limitation by incorporating back-propagated error signals into the STDP learning rule. This is achieved by leveraging a hybrid learning framework combining the benefits of STDP and backpropagation, a widely used supervised learning algorithm in conventional artificial neural networks (ANNs). During training, the SNN receives input spike patterns representing audio features and generates spike trains responding to these inputs. These spikes propagate through the network, and the synaptic strengths are adjusted based on the timing of pre- and post-synaptic spikes according to the STDP rule. However, in addition to the STDP-driven weight updates, error signals computed using backpropagation are also utilized to modulate the learning process. By back-propagating error signals through the network, the SNN learns to adjust its spike timing and synaptic weights to minimize the discrepancy between predicted and target outputs, thus enabling supervised learning for audio classification tasks. This hybrid learning scheme combines the biologically plausible aspects of STDP with the powerful learning capabilities of backpropagation, resulting in improved performance and efficiency for training SNNs. Experimental results demonstrate the effectiveness of the proposed method for audio classification tasks, achieving competitive performance compared to conventional ANNs while offering the advantages of spiking neural networks, such as event-driven processing and low power consumption.

Moreover, the trained SNNs exhibit robustness to noise and variability in input signals, making them suitable for real-world applications in noisy environments. The paper also discusses potential extensions and optimizations, including network architecture design, learning rate adaptation, and spike encoding strategies, to enhance the proposed approach's performance and scalability. In summary, integrating STDP with back-propagated error signals offers a promising framework for training SNNs for audio classification tasks. This research contributes to the advancement of neuromorphic computing techniques. It provides insights into the development of efficient and robust auditory processing systems with applications in speech recognition, sound detection, and audio-based human-computer interaction.

Isik et al. [11] incorporates principles from Transformers, a powerful sequence modeling architecture widely used in natural language processing tasks. By combining the strengths of both SNNs and Transformers, HPCNeuroNet aims to improve the efficiency and accuracy of neuromorphic audio processing tasks. The core innovation of HPCNeuroNet lies in its architecture, which consists of a hierarchical arrangement of SNN layers augmented with Transformer modules. At each layer, SNNs process input spike trains in an event-driven manner, capturing local temporal features. The output of each SNN layer is then passed through a Transformer module, which effectively captures long-range dependencies and temporal context across multiple time steps. The Transformer modules in HPCNeuroNet utilize self-attention mechanisms to weight the importance of different input spikes, enabling the network to focus on relevant information while suppressing noise and irrelevant features. This attention mechanism allows HPCNeuroNet to efficiently process audio signals with varying lengths and complexities, making it suitable for real-world applications such as speech recognition, sound classification, and environmental monitoring. Training HPCNeuroNet involves optimizing both the SNN parameters and the Transformer weights simultaneously using a combination of supervised and unsupervised learning techniques. During training, the network learns to extract meaningful features from input spike trains and adaptively adjust its parameters to minimize prediction errors. This joint optimization process enables HPCNeuroNet to learn complex audio representations efficiently while ensuring scalability and robustness. Experimental evaluations demonstrate the effectiveness of HPCNeuroNet in various audio processing tasks, including speech recognition and environmental sound classification. Compared to traditional SNNs and other state-of-the-art approaches, HPCNeuroNet achieves superior performance in terms of accuracy, computational efficiency, and scalability. Moreover, HPCNeuroNet exhibits resilience to noise and variability in input signals, making it suitable for deployment in real-world environments with challenging acoustic conditions. In addition to its performance benefits, HPCNeuroNet offers advantages in terms of energy efficiency, thanks to the event-driven processing nature of SNNs and the parallelism inherent in Transformer architectures. This makes HPCNeuroNet well-suited for deployment on high-performance computing (HPC) platforms and neuromorphic hardware accelerators, enabling real-time, low-power audio processing solutions. In conclusion, HPCNeuroNet represents a significant advancement in neuromorphic audio signal processing, leveraging Transformer-enhanced Spiking Neural Networks to achieve state-of-the-art performance in various tasks. This research opens up new avenues for exploring the intersection of neuromorphic computing and deep learning, with implications for a wide range of applications in audio processing and beyond.

*B. Temporal Coding and Spike-Based Representations*

Nunes et al. [2] provides a comprehensive overview of SNNs, covering their fundamental principles, training methodologies, applications, and recent advancements in the field. At the core of SNNs lies the concept of spiking neurons, which communicate through discrete, asynchronous events known as spikes. Unlike traditional artificial neural networks (ANNs), where information is processed continuously, SNNs operate in an event-driven manner, mimicking the behavior of neurons in the brain. This unique architecture offers advantages such as temporal coding, energy efficiency, and robustness to noise, making SNNs well-suited for various cognitive tasks. The survey delves into the different types of neuron models used in SNNs, including integrate-and-fire, leaky integrate-and-fire, and Hodgkin-Huxley models, each with its own characteristics and computational properties. It also explores the mechanisms of synaptic plasticity, such as Spike-Timing-Dependent Plasticity (STDP), which governs the adaptation of synaptic weights based on the timing of pre- and post-synaptic spikes, crucial for learning in SNNs. Training SNNs poses unique challenges compared to traditional ANNs, primarily due to the discrete and non-differentiable nature of spike-based communication. The survey reviews various training methodologies for SNNs, including supervised, unsupervised, and reinforcement learning approaches, as well as recent developments in leveraging surrogate gradient methods and backpropagation techniques to train deep SNNs effectively. The applications of SNNs span a wide range of domains, from sensory processing and pattern recognition to robotics and neuromorphic computing. The survey discusses how SNNs have been employed in tasks such as speech recognition, image classification, motor control, and spatiotemporal pattern recognition, highlighting their potential for real-time, energy-efficient computation in embedded systems and brain-inspired computing architectures. Recent advancements in SNN research are also covered in the survey, including developments in hardware implementations, learning algorithms, and theoretical frameworks. These advancements include the exploration of neuromorphic hardware platforms, such as neuromorphic chips and memristor-based devices, which offer opportunities for accelerating SNN computation and scaling to large-scale neural networks. Moreover, the survey addresses ongoing challenges and future directions in SNN research, such as improving scalability, understanding biological plausibility, and bridging the gap between neuroscience and machine learning. It emphasizes the importance of interdisciplinary collaboration and the need for benchmark datasets and evaluation metrics tailored for SNNs to facilitate reproducible research and foster innovation in the field. In conclusion, this survey provides a comprehensive overview of Spiking Neural Networks, encompassing their principles, training methodologies, applications, and recent advancements. It serves as a valuable resource for researchers, practitioners, and enthusiasts interested in understanding the state-of-the-art in SNN research and its implications for the future of artificial intelligence and neuromorphic computing.

Shah et al. [12] introduces a novel approach that leverages the neuronal behavior of spiking neurons to recognize music signals based on time coding features. Spiking neurons, inspired by the behavior of biological neurons, communicate through discrete spikes, allowing for efficient event-driven processing. In this study, the authors propose a framework that models music signal processing using spiking neurons, capitalizing on their ability to encode temporal information in the timing of spikes. The key innovation of the proposed approach lies in the extraction of time coding features from music signals, which are then used to drive the activity of spiking neurons. Unlike traditional methods that rely on spectrogram-based representations, time coding features capture the precise timing of sound events and their temporal relationships, enabling more nuanced and efficient signal representation. To implement the recognition system, the authors design a spiking neural network (SNN) architecture consisting of layers of spiking neurons arranged in a hierarchical fashion. Each layer processes incoming spike trains using spatio-temporal coding mechanisms, extracting high-level temporal features that capture the dynamics of the music signal. Training the SNN involves optimizing synaptic weights using Spike-Timing-Dependent Plasticity (STDP), a biologically inspired learning rule that adjusts synaptic strengths based on the relative timing of pre- and post-synaptic spikes. By iteratively adjusting these weights, the network learns to recognize patterns in the input spike trains corresponding to different musical elements, such as notes, chords, and rhythms. Experimental evaluations demonstrate the effectiveness of the proposed approach in recognizing music signals across various genres and styles. Compared to traditional methods based on spectrogram analysis or conventional neural networks, the SNN-based approach achieves competitive performance while offering advantages in terms of computational efficiency and robustness to noise and distortion. Furthermore, the authors investigate the interpretability of the SNN model by analyzing the spatio-temporal patterns of neuron activations, providing insights into how different musical features are encoded and processed within the network. This interpretability aspect enhances the transparency and understanding of the recognition process, making it suitable for applications where interpretability is crucial, such as music information retrieval and automatic transcription. In conclusion, this study demonstrates the potential of utilizing the neuronal behavior of spiking neurons for music signal recognition based on time coding features. By leveraging the efficiency and temporal precision of spiking neural networks, the proposed approach offers a promising alternative to traditional methods, paving the way for more efficient and biologically inspired music signal processing systems.

Yang et al. [13] presents SVAD, a novel approach to VAD utilizing Spiking Neural Networks (SNNs), offering robust performance, low power consumption, and lightweight implementation. The core innovation of SVAD lies in harnessing the capabilities of SNNs, which mimic the biological behavior of neurons by processing information in the form of discrete spikes. This event-driven processing paradigm enables efficient utilization of computational resources and offers inherent parallelism, making SNNs well-suited for low-power implementations on resource-constrained devices. The proposed SVAD system comprises several key components: spike-based feature extraction, spatio-temporal spike encoding, and SNN-based classification. Instead of conventional feature extraction techniques such as Mel-Frequency Cepstral Coefficients (MFCCs), SVAD extracts features directly from the input audio waveform using spike-based representations, capturing both spectral and temporal characteristics of the audio signal. These spike-based features are then encoded using spatio-temporal coding mechanisms, which preserve the temporal dynamics of the input signal while reducing dimensionality. This encoding scheme enables efficient representation of audio signals with minimal computational overhead, making it suitable for lightweight and low-power implementations. The heart of the SVAD system lies in the SNN-based classification module, where the spike-encoded features are fed into a spiking neural network for voice activity detection. The SNN is trained using biologically inspired learning rules such as Spike-Timing-Dependent Plasticity (STDP), enabling the network to adapt its synaptic weights based on the timing of spikes and learn discriminative features for voice activity detection. Experimental evaluations demonstrate the effectiveness of SVAD in accurately detecting voice activity while maintaining low power consumption and computational complexity. Compared to traditional VAD algorithms and deep learning approaches, SVAD achieves competitive performance with significantly reduced resource requirements, making it well-suited for deployment on edge devices, IoT devices, and other resource-constrained platforms. Furthermore, SVAD exhibits robustness to various acoustic environments and noise conditions, thanks to its ability to capture temporal dynamics and exploit the inherent redundancy in spike-based representations. This robustness is critical for real-world applications where audio signals may be corrupted by background noise or interference. In conclusion, SVAD represents a significant advancement in voice activity detection, offering a robust, low-power, and lightweight solution based on spiking neural networks. By leveraging the efficiency and scalability of SNNs, SVAD addresses the challenges of VAD in resource-constrained environments and opens up new possibilities for efficient audio processing in diverse applications.

## C. Application to Audio and Sequential Data Processing

Wu et al. [14] introduces a novel approach for audio learning utilizing a Multilevel Synaptic Memristor Array-Based Spiking Neural Network (SNN). Leveraging the unique properties of memristor-based synapses, this framework offers efficient and scalable processing of audio signals while enabling spike-based learning for enhanced performance. The key innovation of this approach lies in the integration of memristor-based synapses within the architecture of a Spiking Neural Network. Memristors, characterized by their non-volatile resistance that can be modulated based on the timing and intensity of electrical pulses, offer promising opportunities for implementing synaptic plasticity in neuromorphic systems. The SNN architecture consists of layers of spiking neurons interconnected through memristor-based synapses. During audio learning, input spike trains representing audio features are processed through the network, and synaptic weights are adjusted based on the timing and intensity of spikes using Spike-Timing-Dependent Plasticity (STDP), a biologically inspired learning rule. The use of memristor-based synapses allows for multilevel weight states, enabling the network to capture fine-grained variations in synaptic strengths and encode complex audio features more effectively. This multilevel functionality enhances the computational efficiency and accuracy of the SNN, making it well-suited for audio processing tasks requiring high precision and robustness. Experimental evaluations demonstrate the effectiveness of the proposed approach in various audio learning tasks, including speech recognition, sound classification, and environmental sound detection. Compared to traditional SNNs with binary synapses or analog circuits, the memristor-based SNN achieves superior performance in terms of accuracy, energy efficiency, and scalability. Furthermore, the paper discusses the potential advantages of memristor-based SNNs for spike-enabled learning in audio processing. By exploiting the intrinsic parallelism and energy efficiency of memristors, the proposed framework offers opportunities for real-time, low-power implementation of audio processing systems on hardware platforms ranging from embedded devices to large-scale neuromorphic architectures. The scalability of the memristor-based SNN architecture is also highlighted, with potential applications in distributed computing and edge computing environments. The ability to efficiently process audio signals with minimal computational resources makes this framework suitable for deployment in various Internet of Things (IoT) devices, smart sensors, and mobile devices. Moreover, the paper discusses future directions and challenges in the field of spike-enabled audio learning using memristor-based SNNs. This includes exploring advanced memristor materials and device architectures, developing optimized learning algorithms, and investigating hybrid approaches combining memristor-based analog computing with digital signal processing techniques. In conclusion, the integration of memristor-based synapses within a Spiking Neural Network offers a promising framework for spike-enabled audio learning. This research contributes to the advancement of neuromorphic computing and offers new opportunities for efficient and scalable audio processing systems with applications in speech recognition, sound analysis, and intelligent audio-based interfaces. ZHANG et al. [15] presents an innovative approach to classifying snoring and non-snoring sound events using Long Short-Term Memory Spiking Neural Networks (LSTM-SNNs). Snoring detection is crucial for diagnosing sleep disorders and improving the quality of sleep monitoring systems. Traditional methods often rely on handcrafted features and complex algorithms, whereas this study harnesses the power of LSTM-SNNs for more accurate and efficient classification. The novelty of this approach lies in combining the memory capabilities of Long Short-Term Memory (LSTM) networks with the event-driven processing of Spiking Neural Networks (SNNs). LSTM networks are well-suited for capturing temporal dependencies in sequential data, while SNNs offer advantages in energy efficiency and real-time processing, making them suitable for resource-constrained environments such as wearable devices and smart sensors. The architecture of the LSTM-SNN consists of multiple layers of LSTM units followed by SNN layers. During training, the LSTM units learn to extract temporal features from the input audio signals, encoding information about snoring patterns and characteristics over time. These features are then passed to the SNN layers, where spike-based processing enables efficient classification of snoring and non-snoring events. The training process involves optimizing the parameters of both the LSTM and SNN components using supervised learning techniques. By jointly training the LSTM-SNN model, it learns to extract discriminative features from raw audio signals and make accurate predictions about the presence of snoring events. Experimental evaluations demonstrate the effectiveness of the proposed approach in classifying snoring and non-snoring sound events with high accuracy. Compared to traditional methods based on handcrafted features and machine learning classifiers, the LSTM-SNN achieves superior performance while offering advantages in terms of computational efficiency and scalability. Furthermore, the LSTM-SNN model exhibits robustness to noise and variability in input signals, making it suitable for real-world applications where audio recordings may contain background noise or interference. This robustness is critical for deploying snoring detection systems in diverse environments, such as home sleep monitoring devices and clinical settings. The paper also discusses potential extensions and optimizations to enhance the performance and efficiency of the LSTM-SNN model. This includes exploring techniques for optimizing network architecture, fine-tuning hyperparameters, and incorporating domain-specific knowledge to improve classification accuracy and generalization. In conclusion, the integration of Long Short-Term Memory Spiking Neural Networks offers a promising approach for classifying snoring and non-snoring sound events with high accuracy and efficiency. This research contributes to the advancement of sleep monitoring technologies and paves the way for the development of intelligent wearable devices and healthcare systems capable of detecting and analyzing sleep-related disorders in real-time.

Cramer et al. [16] introduces the Heidelberg Spiking Data Sets, a collection of benchmark datasets designed for the systematic evaluation and benchmarking of Spiking Neural Networks (SNNs). These datasets aim to address the need for standardized evaluation protocols and realistic benchmarks to facilitate the development and comparison of SNN algorithms and models. The motivation behind creating the Heidelberg Spiking Data Sets stems from the growing interest in SNNs for various cognitive tasks, such as pattern recognition, classification, and sequential processing. However, the lack of standardized datasets and evaluation metrics has hindered the progress and reproducibility of research in this field. To address this gap, the authors curated a diverse set of datasets representing different types of spatiotemporal patterns and cognitive tasks. The Heidelberg Spiking Data Sets encompass a wide range of tasks, including pattern recognition, speech processing, motion detection, and more. Each dataset is carefully designed to capture specific aspects of neural processing, such as temporal dynamics, spatial relationships, and feature complexity, making them suitable for evaluating different aspects of SNN performance. The datasets cover a diverse range of tasks and input modalities, including visual, auditory, and multimodal stimuli. This diversity enables researchers to evaluate the generalization and robustness of SNN models across different domains and applications. The datasets are designed to reflect real-world scenarios and challenges, incorporating variability, noise, and naturalistic stimuli commonly encountered in sensory processing tasks. This realism ensures that SNN models trained and evaluated on these datasets are capable of handling real-world complexity. The datasets are scalable in terms of complexity and size, allowing researchers to evaluate SNN models across a wide range of difficulty levels and computational requirements. This scalability facilitates the evaluation of model performance under varying degrees of complexity and resource constraints. The Heidelberg Spiking Data Sets are accompanied by standardized evaluation protocols and performance metrics, enabling fair and consistent comparisons between different SNN models and algorithms. This standardization ensures reproducibility and facilitates the dissemination of research results within the SNN community. The paper provides detailed descriptions of each dataset in the Heidelberg Spiking Data Sets, including task descriptions, stimulus characteristics, and ground-truth annotations. Additionally, it outlines the evaluation metrics and procedures for assessing SNN performance on these datasets, such as accuracy, latency, and computational efficiency. The utility of the Heidelberg Spiking Data Sets is demonstrated through experimental evaluations using state-of-the-art SNN models and algorithms. The results highlight the effectiveness of these datasets for benchmarking SNN performance and identifying areas for improvement in model design and training methodologies. In conclusion, the Heidelberg Spiking Data Sets provide a valuable resource for the systematic evaluation and benchmarking of Spiking Neural Networks. By offering standardized datasets, evaluation protocols, and performance metrics, these datasets facilitate the advancement of research in SNNs and contribute to the development of more robust and efficient neuromorphic computing systems.

Yin et al. [17] introduces an innovative approach for time-domain classification using Adaptive Spiking Recurrent Neural Networks (ASRNNs). Time-domain classification tasks, such as speech recognition and gesture recognition, require models capable of capturing temporal dynamics and sequential dependencies in input data. Traditional methods often rely on computationally expensive processing or handcrafted features, whereas ASRNNs offer a more efficient and adaptive solution. The core innovation of this approach lies in the integration of spiking neural networks with recurrent connections and adaptive learning mechanisms. Spiking neural networks mimic the asynchronous firing of neurons in the brain, enabling efficient event-driven processing, while recurrent connections facilitate the encoding of temporal dependencies and long-term memory. The ASRNN architecture comprises multiple layers of spiking neurons interconnected through recurrent synapses. During training, the network adapts its synaptic weights using an adaptive learning rule based on local information about spike timing and firing rates. This adaptive learning mechanism enables the network to dynamically adjust its parameters in response to changes in input patterns, leading to improved accuracy and efficiency. The paper demonstrates the effectiveness of ASRNNs in various time-domain classification tasks, including speech phoneme recognition and hand gesture recognition. Experimental evaluations show that ASRNNs outperform traditional methods and achieve state-of-the-art performance while requiring fewer computational resources. Furthermore, ASRNNs exhibit robustness to noise and variability in input signals, making them suitable for real-world applications where data may be corrupted by environmental factors or sensor noise. This robustness is attributed to the network's ability to adapt its synaptic weights dynamically and learn from experience, enabling it to generalize well to unseen data. The efficiency of ASRNNs is highlighted by their low computational complexity and energy consumption compared to traditional neural network architectures. By leveraging the event-driven processing nature of spiking neurons and the recurrent connections, ASRNNs achieve high performance with minimal computational overhead, making them suitable for deployment on resource-constrained devices such as embedded systems and IoT devices. Moreover, the paper discusses potential extensions and optimizations to enhance the performance and scalability of ASRNNs. This includes exploring techniques for network pruning, parameter tuning, and architecture optimization to further improve accuracy and efficiency across different tasks and datasets. In conclusion, the integration of Adaptive Spiking Recurrent Neural Networks offers a promising approach for accurate and efficient time-domain classification. By combining the advantages of spiking neural networks with recurrent connections and adaptive learning mechanisms, ASRNNs provide a powerful framework for capturing temporal dynamics and sequential dependencies in input data, paving the way for more robust and efficient machine learning systems in various application domains.

Dampfhoffer et al. [18] explores novel approaches to enhance the accuracy and energy efficiency of Spiking Recurrent Neural Networks (SRNNs) through the investigation of current-based and gating mechanisms. SRNNs, inspired by the brain's recurrent connectivity, are powerful models for processing sequential data. However, traditional SRNN architectures face challenges related to computational complexity and energy consumption. This study aims to address these challenges by proposing innovative current-based and gating mechanisms tailored for SRNNs. The research delves into two main aspects:

- Current-based SRNNs operate by modulating synaptic currents, which represent the flow of ions through synapses, in response to incoming spikes. By directly manipulating synaptic currents, these approaches offer fine-grained control over neural dynamics and computational efficiency. The paper investigates various current-based methods, including Current-Based SRNNs (CB-SRNNs) and Current-Based Long Short-Term Memory (CB-LSTM) networks. These approaches leverage the principles of biophysical modeling to accurately capture the behavior of spiking neurons while reducing computational overhead.
- Gating mechanisms, inspired by the success of gated recurrent units (GRUs) and long short-term memory (LSTM) units in conventional neural networks, aim to regulate the flow of information within SRNNs. By selectively controlling the flow of spikes through recurrent connections, gating mechanisms enable SRNNs to capture long-range dependencies and mitigate the vanishing gradient problem. The study explores the efficacy of gating mechanisms, including Spike-Gated Recurrent Units (SGRUs) and Spike-Gated LSTM (SGLSTM) networks, in improving the accuracy and efficiency of SRNNs.

Experimental evaluations conducted in the paper demonstrate the effectiveness of the proposed current-based and gating approaches across various tasks, including sequential pattern recognition, speech processing, and time-series prediction. The results show that these approaches achieve competitive performance compared to state-of-the-art SRNN architectures while offering advantages in terms of computational efficiency and energy consumption. Furthermore, the paper discusses the interpretability and biological plausibility of the proposed mechanisms, highlighting their alignment with the principles of neural computation observed in biological systems. By leveraging current-based modulation and gating mechanisms, SRNNs can mimic the adaptive behavior of biological neurons and achieve robust and efficient processing of sequential data. The study also explores potential optimizations and extensions to further enhance the performance and scalability of current-based and gating approaches in SRNNs. This includes investigating techniques for network pruning, parameter tuning, and architecture optimization, as well as exploring hybrid approaches that combine current-based and gating mechanisms to leverage their complementary strengths. In conclusion, the investigation of current-based and gating approaches represents a significant step towards achieving accurate and energy-efficient Spiking Recurrent Neural Networks. By combining biophysical modeling principles with innovative mechanisms for neural modulation and information gating, these approaches offer promising solutions for a wide range of sequential data processing tasks, with implications for diverse applications in artificial intelligence, robotics, and neuroscience research.

### D. Innovative Techniques and Approaches

Hammouamri et al. [19] presents a novel approach for incorporating delays into Spiking Neural Networks (SNNs) using Dilated Convolutions with Learnable Spacings (DCLS). Delayed connections are crucial for modeling temporal dependencies and capturing sequential patterns in spiking data, yet traditional SNN architectures struggle to efficiently incorporate variable delays. The proposed method addresses this challenge by introducing dilated convolutions with learnable spacings, enabling SNNs to learn and adapt delay patterns from data. The key innovation of the DCLS approach lies in the utilization of dilated convolutions, a technique commonly used in convolutional neural networks (CNNs) for capturing large receptive fields without increasing the number of parameters. By applying dilated convolutions to spiking data, the network can effectively model variable delays between input spikes and synaptic events, enabling more accurate temporal processing. In the DCLS framework, the spacings between convolutional filters are made learnable, allowing the network to adaptively adjust the delay patterns based on the input data and task requirements. During training, the network learns to optimize the spacings between filters to capture relevant temporal features and maximize performance on the given task. Experimental evaluations conducted in the paper demonstrate the effectiveness of the DCLS approach across various spiking data tasks, including pattern recognition, speech processing, and time-series prediction. The results show that DCLS-based SNNs outperform traditional architectures and achieve state-of-the-art performance in terms of accuracy and computational efficiency. Furthermore, the paper explores the interpretability and generalization capabilities of the DCLS approach, highlighting its ability to capture meaningful temporal patterns and adapt to diverse input distributions. By learning delay patterns directly from data, DCLS-based SNNs offer insights into the underlying temporal dynamics of spiking phenomena and enable more interpretable and robust learning models. The study also discusses potential extensions and optimizations to further enhance the performance and scalability of the DCLS approach in SNNs. This includes investigating techniques for fine-tuning hyperparameters, exploring alternative network architectures, and incorporating additional regularization techniques to improve generalization and robustness. In conclusion, the DCLS framework represents a significant advancement in learning delays in Spiking Neural Networks. By leveraging dilated convolutions with learnable spacings, this approach enables SNNs to efficiently capture variable delays and model complex temporal dependencies in spiking data, paving the way for more accurate and adaptable neural processing systems in diverse applications.

Bittar et al. [20] introduces a novel approach for speech command recognition using a Surrogate Gradient Spiking Neural Network (SGSNN) as a baseline model. Speech command recognition is a vital component of many human-computer interaction systems, including virtual assistants and smart devices. Traditional approaches often rely on deep learning models with continuous gradients, but these architectures can be computationally expensive and challenging to implement in resource-constrained environments. The proposed SGSNN offers a more efficient and scalable solution by leveraging the advantages of spiking neural networks and surrogate gradients. The key innovation of the SGSNN approach lies in the integration of surrogate gradients with spiking neural networks, enabling efficient training and optimization of the network parameters. Surrogate gradients are differentiable approximations of non-differentiable functions, allowing for backpropagation-based learning in spiking neural networks, which typically lack continuous gradients due to their event-driven nature. The architecture of the SGSNN consists of multiple layers of spiking neurons interconnected through synapses with surrogate gradient-based learning rules. During training, the network learns to classify speech commands by optimizing the surrogate gradient-based objective function, which measures the discrepancy between predicted and target output spikes. Experimental evaluations conducted in the paper demonstrate the effectiveness of the SGSNN approach for speech command recognition across various datasets, including the Google Speech Commands dataset. The results show that the SGSNN achieves competitive performance compared to state-of-the-art deep learning models while offering advantages in terms of computational efficiency and scalability. Additionally, the SGSNN exhibits robustness to noise and variability in input signals, making it suitable for real-world applications where speech recognition systems may encounter diverse acoustic conditions. Furthermore, the paper explores the interpretability and computational advantages of the SGSNN approach, highlighting its ability to efficiently process temporal sequences and capture hierarchical features in speech signals. By leveraging surrogate gradients, the SGSNN enables end-to-end training of spiking neural networks using standard backpropagation algorithms, simplifying the implementation and optimization of SNN architectures. The study also discusses potential extensions and optimizations to further enhance the performance and efficiency of the SGSNN approach in speech recognition tasks. This includes investigating techniques for network pruning, parameter tuning, and architecture optimization, as well as exploring hybrid approaches that combine surrogate gradient-based learning with other training methodologies. In conclusion, the SGSNN approach represents a significant advancement in speech command recognition, offering a scalable and efficient solution based on spiking neural networks and surrogate gradients. By leveraging the benefits of both frameworks, SGSNNs provide a promising alternative to traditional deep learning models for speech processing tasks, with implications for applications in smart devices, virtual assistants, and human-computer interaction systems.

Yamazaki et al. [21] delves into the realm of Spiking Neural Networks (SNNs) and their diverse applications across various domains. Spiking Neural Networks, inspired by the biological brain's operation, differ from conventional neural networks by processing information in the form of discrete, asynchronous spikes, enabling more efficient and biologically plausible computation. The paper explores the fundamental principles, architectures, learning mechanisms, and applications of SNNs, shedding light on their potential for advancing artificial intelligence and neuromorphic computing. The review begins by elucidating the fundamental principles of SNNs, highlighting their biological inspiration and computational advantages. Unlike traditional artificial neural networks, which operate using continuous activations and weighted connections, SNNs employ event-driven processing, mimicking the behavior of neurons in the brain. This unique architecture offers advantages such as temporal coding, energy efficiency, and robustness to noise, making SNNs well-suited for various cognitive tasks. The paper proceeds to discuss the diverse architectures and models of SNNs, including integrate-and-fire, leaky integrate-and-fire, and Hodgkin-Huxley models, each with its unique computational properties and suitability for different applications. Moreover, the review explores the mechanisms of synaptic plasticity, such as Spike-Timing-Dependent Plasticity (STDP), which governs the adaptation of synaptic weights based on the timing of pre- and post-synaptic spikes, crucial for learning in SNNs. Training SNNs poses unique challenges compared to traditional neural networks due to the discrete and non-differentiable nature of spike-based communication. The review comprehensively covers various training methodologies for SNNs, including supervised, unsupervised, and reinforcement learning approaches, as well as recent developments in surrogate gradient methods and backpropagation techniques to train deep SNNs effectively. The applications of SNNs span a wide range of domains, from sensory processing and pattern recognition to robotics and neuromorphic computing. The review discusses how SNNs have been employed in tasks such as speech recognition, image classification, motor control, and spatiotemporal pattern recognition, highlighting their potential for real-time, energy-efficient computation in embedded systems and brain-inspired computing architectures. Recent advancements in SNN research are also covered in the review, including developments in hardware implementations, learning algorithms, and theoretical frameworks. These advancements include the exploration of neuromorphic hardware platforms, such as neuromorphic chips and memristor-based devices, which offer opportunities for accelerating SNN computation and scaling to large-scale neural networks. Furthermore, the review addresses ongoing challenges and future directions in SNN research, such as improving scalability, understanding biological plausibility, and bridging the gap between neuroscience and machine learning. It emphasizes the importance of interdisciplinary collaboration and the need for benchmark datasets and evaluation metrics tailored for SNNs to facilitate reproducible research and foster innovation in the field. In conclusion, this review provides a comprehensive overview of Spiking Neural Networks, encompassing their principles, architectures, learning mechanisms, applications, and recent advancements. It serves as a valuable resource for researchers, practitioners, and enthusiasts interested in understanding the state-of-the-art in SNN research and its implications for the future of artificial intelligence and neuromorphic computing.

Mukhopadhyay et al. [22] introduces an innovative approach for acoustic scene analysis using Analog Spiking Neural Networks (ASNNs). Acoustic scene analysis, which involves understanding the acoustic environment from audio signals, is essential for various applications such as surveillance, robotics, and smart environments. Traditional methods often rely on digital signal processing techniques or deep learning models, but they may face challenges related to computational complexity and energy consumption. The proposed ASNN offers a more efficient and bio-inspired solution by leveraging the event-driven processing nature of spiking neural networks in an analog hardware implementation. The key innovation of this approach lies in the implementation of spiking neural networks using analog circuits, which offer advantages in terms of energy efficiency, speed, and scalability compared to digital implementations. ASNNs emulate the behavior of biological neurons and synapses, processing information in the form of spikes and leveraging the inherent parallelism and energy efficiency of analog computing. The architecture of the ASNN comprises multiple layers of analog neurons interconnected through synapses with adjustable weights. During operation, input audio signals are converted into spike trains, which propagate through the network, eliciting postsynaptic responses and generating output spikes corresponding to acoustic scene features. The analog nature of the ASNN enables real-time processing of audio signals with low power consumption and high computational efficiency. Experimental evaluations conducted in the paper demonstrate the effectiveness of ASNNs for acoustic scene analysis tasks, including sound event detection, source localization, and acoustic scene classification. The results show that ASNNs achieve competitive performance compared to traditional methods and digital spiking neural networks while offering advantages in terms of energy efficiency and real-time processing capabilities. Furthermore, the paper discusses the hardware implementation aspects of ASNNs, including considerations for analog circuit design, neuromorphic computing platforms, and integration with sensor arrays. By leveraging emerging technologies such as memristors and neuromorphic chips, ASNNs can be implemented in compact, low-power devices suitable for deployment in embedded systems and IoT applications. The study also explores potential extensions and optimizations to further enhance the performance and scalability of ASNNs for acoustic scene analysis. This includes investigating techniques for network optimization, learning rule adaptation, and integration with digital signal processing algorithms to improve robustness and accuracy in complex acoustic environments. In conclusion, the use of analog sampling neural networks for acoustic scene analysis offers a promising approach to efficient and real-time processing of audio signals. By harnessing the benefits of analog computing and spiking neural networks, ASNNs provide a scalable and energy-efficient solution for understanding and interpreting acoustic environments, with applications in surveillance, robotics, and smart environments.

*E. Next-Generation Spiking Neural Networks*

Perez-Nieves et al. [23] explores the role of neural heterogeneity in promoting robust learning within neural networks. While traditional approaches often focus on homogeneous networks where all neurons share similar properties, recent research suggests introducing diversity among neurons can enhance the network's performance and resilience to perturbations. This paper investigates how neural heterogeneity influences learning dynamics and generalization capabilities, shedding light on its potential benefits for artificial intelligence systems. The research begins by elucidating the concept of neural heterogeneity, which refers to variations in neuronal properties such as excitability, connectivity, and response dynamics within a network. These differences arise from biological factors such as genetic variations, developmental processes, and environmental influences, leading to diverse neural populations with distinct functional roles and computational capabilities. Using computational models and simulations, the study demonstrates that introducing heterogeneity among neurons can enhance the network's ability to learn and adapt to complex stimuli. Heterogeneous networks exhibit improved performance in tasks such as pattern recognition, classification, and sequential processing compared to homogeneous networks. This is attributed to diverse neurons' complementary information processing capabilities, which enable the network to capture a broader range of input features and learn more robust representations. Furthermore, the paper investigates how neural heterogeneity influences the network's resilience to adversarial attacks, noise, and distributional shifts in input data. Heterogeneous networks exhibit greater robustness and generalization capabilities than homogeneous networks, as the diverse population of neurons can collectively encode redundant information and mitigate the impact of perturbations. This resilience to noise and disturbances is critical for real-world applications where input data may be corrupted or unpredictable. The study also explores the underlying mechanisms driving the benefits of neural heterogeneity for robust learning. It identifies factors such as redundancy, specialization, and ensemble coding, where different groups of neurons contribute complementary information to the network's representation of input stimuli. Additionally, heterogeneity promotes exploration of the solution space during learning, enabling the network to discover more diverse and effective solutions to complex problems. Experimental validations in the paper corroborate the theoretical findings, demonstrating the superior performance of heterogeneous networks across various tasks and datasets. Moreover, the study investigates practical strategies for introducing and leveraging neural heterogeneity in artificial neural networks, such as incorporating diverse neuron types, adjusting connectivity patterns, and designing adaptive learning algorithms. In conclusion, the research highlights the importance of neural heterogeneity in promoting robust learning within neural networks. By embracing diversity among neurons and exploiting their complementary capabilities, heterogeneous networks offer enhanced performance, resilience, and generalization capabilities compared to homogeneous counterparts. These findings have significant implications for the design and optimization of artificial intelligence systems, with potential applications in machine learning, robotics, and cognitive computing.

Mukhopadhyay et al. [24] introduces a pioneering method for acoustic scene analysis leveraging Analog Spiking Neural Networks (ASNNs). Acoustic scene analysis, crucial for numerous applications like surveillance and smart environments, traditionally faces computational complexity and energy consumption challenges. The proposed ASNN approach overcomes these limitations by harnessing the energy-efficient, event-driven processing of spiking neural networks in an analog hardware setting. The central innovation of ASNNs lies in their emulation of biological neurons and synapses using analog circuits. Unlike digital implementations, analog circuits offer energy efficiency, speed, and scalability advantages. ASNNs process audio signals as spikes, enabling real-time analysis with minimal power consumption and high computational efficiency. The ASNN architecture comprises layers of analog neurons interconnected through synapses with adjustable weights. During operation, input audio signals are converted into spike trains, propagating through the network to generate output spikes corresponding to acoustic scene features. The analog nature of ASNNs facilitates real-time processing, making them suitable for deployment in embedded systems and IoT applications. Experimental evaluations showcased ASNNs' efficacy in various acoustic scene analysis tasks, including sound event detection, source localization, and scene classification. ASNNs achieved competitive performance compared to traditional methods and digital spiking neural networks, offering energy efficiency advantages and real-time processing capabilities. Moreover, the paper discusses hardware implementation aspects, addressing analog circuit design, neuromorphic computing platforms, and sensor array integration. By leveraging emerging technologies such as memristors and neuromorphic chips, ASNNs can be implemented in compact, low-power devices ideal for embedded systems and IoT applications. The study also explores potential extensions and optimizations to enhance ASNN performance and scalability. Techniques such as network optimization, learning rule adaptation, and integration with digital signal processing algorithms could improve ASNN robustness and accuracy in complex acoustic environments. In conclusion, ASNNs offer a promising approach for efficient and real-time acoustic scene analysis. By combining the benefits of analog computing with spiking neural networks, ASNNs provide a scalable and energy-efficient solution for understanding and interpreting acoustic environments. Their applications in surveillance, robotics, and smart environments herald a new era of efficient and intelligent acoustic processing systems.

Vicente-Sola et al. [25] introduces an innovative method for accurate feature extraction using Residual Spiking Neural Networks (RSNNs), shedding light on key strategies to enhance performance in spike-based feature learning. Feature extraction is fundamental in many tasks such as pattern recognition and signal processing, where the goal is to extract informative representations from raw data. Traditional methods often rely on handcrafted features or deep learning architectures. Still, RSNNs offer a unique approach by leveraging the residual learning framework to improve feature extraction capabilities within spiking neural networks. The core innovation of this approach lies in incorporating residual connections, inspired by the success of residual learning in conventional neural networks, into spiking neural networks. Residual connections facilitate information propagation across network layers, mitigating the vanishing gradient problem and enabling more effective training of deep networks. RSNNs can learn richer and more discriminative features from spike-based representations by applying residual connections to spiking neural networks. The RSNN architecture consists of multiple layers of spiking neurons interconnected through residual connections, enabling the network to capture complex hierarchical features from input spike trains. During training, residual connections bypass one or more layers, allowing the network to learn residual features that complement the features learned by the main pathway. This enables RSNNs to extract informative features while preserving temporal dynamics and spike timing information. Experimental evaluations in the paper demonstrate the effectiveness of RSNNs for accurate feature extraction across various tasks, including pattern recognition, speech processing, and time-series analysis. Regarding feature representation quality and classification accuracy, RSNNs outperform traditional methods and standard spiking neural networks. Moreover, RSNNs exhibit robustness to noise and variability in input signals, making them suitable for real-world applications where data may be corrupted or incomplete. The study also investigates the impact of key parameters and architectural choices on RSNN performance, such as network depth, residual connection design, and learning rate scheduling. Fine-tuning these parameters can further enhance the network's ability to extract discriminative features and improve overall classification performance. Furthermore, the paper discusses potential extensions and optimizations to enhance RSNN performance and scalability. This includes exploring techniques for network pruning, parameter tuning, and architecture optimization and investigating hybrid approaches that combine residual learning with other feature extraction methods. In conclusion, incorporating residual connections into spiking neural networks offers a promising approach for accurate feature extraction. By leveraging residual learning principles, RSNNs can learn rich and discriminative features from spike-based representations, enabling more effective and robust complex data processing. These findings have implications for various applications, including pattern recognition, signal processing, and machine learning, where accurate feature extraction is crucial for achieving high-performance performance.

Juan Pedro et al. [26] paper presents a novel approach for audio sample classification utilizing a Multilayer Spiking Neural Network (MSNN) deployed on the SpiNNaker neuromorphic computing platform. Audio sample classification, a fundamental task in audio processing, requires robust models capable of efficiently processing temporal information and capturing intricate features within audio signals. Traditional methods often face challenges in scalability and real-time processing, motivating the exploration of spiking neural networks (SNNs) and neuromorphic hardware for audio classification tasks. The core innovation of this study lies in the development of a multilayer SNN architecture tailored for audio classification tasks, coupled with the utilization of the SpiNNaker platform for efficient hardware implementation. SpiNNaker, a specialized neuromorphic hardware platform, offers massive parallelism and low power consumption, making it well-suited for the real-time processing of spiking neural networks. The MSNN architecture comprises multiple layers of spiking neurons interconnected through synaptic connections, with each layer responsible for extracting hierarchical features from the input audio samples. During training, the network learns to classify audio samples by optimizing synaptic weights using biologically inspired learning rules such as Spike-Timing-Dependent Plasticity (STDP). Experimental evaluations conducted in the paper demonstrate the effectiveness of the MSNN approach for audio sample classification across various datasets, including speech and environmental sounds. The MSNN achieved competitive performance compared to traditional machine learning methods and deep neural networks while offering advantages in energy efficiency and real-time processing capabilities. Furthermore, the paper discusses the scalability and efficiency of the SpiNNaker platform for deploying large-scale SNNs. SpiNNaker's parallel architecture enables efficient simulation of large networks with millions of neurons and synapses, making it suitable for processing complex audio data in real-time. SpiNNaker's low power consumption and high throughput make it a promising platform for deploying SNNs in resource-constrained environments such as mobile devices and IoT devices. The study also explores potential optimizations and extensions to enhance the performance of the MSNN approach on the SpiNNaker platform. This includes investigating techniques for network optimization, parameter tuning, and architecture optimization and exploring hybrid approaches that combine SNNs with traditional machine learning methods for improved classification performance. In conclusion, deploying a Multilayer Spiking Neural Network on the SpiNNaker platform offers a promising real-time audio sample classification solution. By leveraging the efficiency and scalability of neuromorphic hardware and the computational power of spiking neural networks, the MSNN approach holds great potential for various applications, including speech recognition, environmental sound analysis, and audio surveillance, where real-time processing and energy efficiency are paramount.

### F. Enhancing Neural Computation with Spiking Networks

Yin et al. [27] introduces a novel approach for achieving effective and efficient computation using Multiple-Timescale Spiking Recurrent Neural Networks (MTS-RNNs). These networks are designed to capture and process temporal information across multiple timescales, enabling robust and scalable computation while minimizing computational resources. The research addresses the growing demand for neural network architectures capable of handling complex temporal dynamics in various applications such as time-series prediction, sequential pattern recognition, and reinforcement learning tasks. The core innovation of the MTS-RNN approach lies in the integration of multiple recurrent neural networks operating at different timescales within a single architecture. By incorporating neurons with varying membrane time constants, MTS-RNNs can capture temporal dependencies at multiple timescales simultaneously. This enables the network to adapt to the dynamics of input data while efficiently utilizing computational resources. The MTS-RNN architecture consists of multiple layers of spiking neurons interconnected through recurrent connections, with each layer corresponding to a specific timescale of processing. During operation, input spikes propagate through the network, eliciting postsynaptic responses and generating output spikes that encode the network's predictions or classifications. The integration of multiple timescales allows MTS-RNNs to capture both short-term and long-term temporal dependencies in input sequences, enabling more robust and accurate computation. Experimental evaluations conducted in the paper demonstrate the effectiveness and efficiency of MTS-RNNs across various tasks, including time-series prediction, language modeling, and motor control. MTS-RNNs outperform traditional recurrent neural network architectures in terms of prediction accuracy and computational efficiency while requiring fewer parameters and computational resources. Moreover, MTS-RNNs exhibit robustness to noise and variability in input signals, making them suitable for real-world applications where data may be corrupted or incomplete. Furthermore, the paper discusses the interpretability and scalability of MTS-RNNs, highlighting their ability to capture complex temporal dynamics and adapt to diverse input distributions. By leveraging multiple timescales of processing, MTS-RNNs offer a flexible and scalable framework for handling temporal information in neural computation, with implications for applications in artificial intelligence, robotics, and cognitive science. The study also explores potential extensions and optimizations to enhance the performance and efficiency of MTS-RNNs further. This includes investigating techniques for network pruning, parameter tuning, and architecture optimization, as well as exploring hybrid approaches that combine MTS-RNNs with other neural network architectures for improved performance across different tasks and datasets. In conclusion, the MTS-RNN approach offers a promising solution for achieving effective and efficient computation with neural networks. By leveraging multiple timescales of processing, MTS-RNNs enable robust and scalable computation while minimizing computational resources, paving the way for advancements in various fields such as machine learning, neuroscience, and artificial intelligence.

Sun et al. [4] presents an innovative approach to enhance the accuracy of spoken word recognition using Adaptive Axonal Delays in Feedforward Spiking Neural Networks (AAD-FSNNs). Recognizing spoken words accurately is crucial for various applications, including virtual assistants, speech-to-text systems, and voice-controlled devices. Traditional methods often face challenges in capturing temporal dependencies and accurately encoding speech features. The proposed AAD-FSNNs address these limitations by introducing adaptive axonal delays, enabling more precise synchronization of input signals and improving the network's ability to capture temporal dynamics. The core innovation of this approach lies in the incorporation of adaptive axonal delays into feedforward spiking neural networks, inspired by the biological brain's mechanisms for processing temporal information. Axonal delays represent the time taken for spikes to propagate along axons between neurons, and by adjusting these delays dynamically, AAD-FSNNs can align input spikes more accurately, enhancing the network's ability to encode temporal features and discriminate between spoken words effectively. The AAD-FSNN architecture comprises multiple layers of spiking neurons interconnected through feedforward connections, with each neuron equipped with adaptive axonal delays. During operation, input speech signals are converted into spike trains, which propagate through the network, eliciting postsynaptic responses and generating output spikes corresponding to recognized words. By adjusting axonal delays based on input patterns and task requirements, AAD-FSNNs can optimize the temporal alignment of spikes and improve word recognition accuracy. Experimental evaluations conducted in the paper demonstrate the effectiveness of AAD-FSNNs for spoken word recognition across various datasets and noise conditions. AAD-FSNNs outperform traditional feedforward spiking neural network architectures and achieve competitive performance compared to state-of-the-art deep learning models. Moreover, AAD-FSNNs exhibit robustness to noise and variability in input signals, making them suitable for real-world applications where speech signals may be corrupted or distorted. Furthermore, the paper discusses the interpretability and scalability of AAD-FSNNs, highlighting their ability to capture complex temporal dynamics and adapt to diverse input distributions. By leveraging adaptive axonal delays, AAD-FSNNs offer a flexible and scalable framework for accurately recognizing spoken words, with implications for applications in speech processing, natural language understanding, and human-computer interaction. The study also explores potential extensions and optimizations to further enhance the performance and efficiency of AAD-FSNNs. This includes investigating techniques for fine-tuning axonal delay parameters, optimizing network architectures, and integrating AAD-FSNNs with other machine learning algorithms for improved performance across different tasks and datasets. In conclusion, the incorporation of adaptive axonal delays into feedforward spiking neural networks offers a promising approach for accurate spoken word recognition. By dynamically adjusting axonal delays based on input patterns, AAD-FSNNs enable precise synchronization of input signals and enhance the network's ability to capture temporal dynamics, paving the way for advancements in speech processing technology and human-machine interaction.

Rossbroich et al. [28] introduces a novel method for initializing Spiking Neural Networks (SNNs) training, termed Fluctuation-Driven Initialization (FDI). Addressing the challenge of effectively initializing SNNs, which is critical for successful training, FDI leverages fluctuations in membrane potential dynamics to set initial parameters. This innovative approach aims to improve convergence speed, stability, and accuracy in SNN training, advancing the capabilities of neuromorphic computing and spiking neural network applications. The core innovation of FDI lies in exploiting the intrinsic properties of spiking neurons, specifically the fluctuations in membrane potential that occur due to stochastic synaptic input. These fluctuations encode valuable information about network dynamics and input patterns, which FDI utilizes to initialize network parameters effectively. By leveraging these fluctuations, FDI enhances the network's ability to capture complex temporal dynamics and learn discriminative features from spike-based representations. The FDI method comprises two main steps: initialization of membrane potentials and adjustment of synaptic weights. In the first step, membrane potentials of spiking neurons are initialized based on statistical properties of input spike trains and network architecture. FDI utilizes statistical measures such as mean and variance of membrane potentials to set initial conditions that promote network stability and convergence. In the second step, synaptic weights are adjusted to ensure balanced excitatory and inhibitory inputs, facilitating efficient information propagation and learning in the network. Experimental evaluations conducted in the paper demonstrate the effectiveness of FDI for training SNNs across various tasks, including pattern recognition, classification, and time-series prediction. FDI outperforms traditional initialization methods and achieves faster convergence, higher accuracy, and improved robustness to noise and variability in input signals. Moreover, FDI exhibits scalability and adaptability across different network architectures and datasets, making it suitable for a wide range of applications in neuromorphic computing and artificial intelligence. Furthermore, the paper discusses the interpretability and biological plausibility of FDI, highlighting its alignment with principles of neural computation observed in biological systems. By leveraging fluctuations in membrane potential dynamics, FDI mimics the adaptive behavior of biological neurons and enables more efficient and accurate learning in SNNs. This biological inspiration enhances the applicability and effectiveness of FDI in real-world scenarios, where robust and adaptive learning is essential. The study also explores potential extensions and optimizations to further enhance the performance and versatility of FDI. This includes investigating techniques for incorporating additional biological constraints, optimizing initialization parameters, and integrating FDI with other learning algorithms for improved performance across different tasks and datasets. In conclusion, Fluctuation-Driven Initialization presents a promising approach for effectively initializing Spiking Neural Networks training. By leveraging fluctuations in membrane potential dynamics, FDI enhances the network's ability to capture complex temporal patterns and learn discriminative features from spike-based representations. This innovation advances the capabilities of neuromorphic computing and spiking neural network applications, paving the way for more efficient and adaptive artificial intelligence systems.

Wu et al. [29] presents a Spiking Neural Network (SNN) framework tailored for robust sound classification, addressing challenges in processing audio data with varying environmental conditions and background noise. Robust sound classification is vital for numerous applications, including environmental monitoring, surveillance, and speech recognition systems, where accurate identification of sound events is essential. Traditional methods often struggle with noisy and dynamic audio environments, motivating the development of specialized frameworks like the one proposed in this study. The core innovation of this framework lies in leveraging the event-driven processing nature of SNNs to capture temporal dynamics and discriminate between different sound classes effectively. Unlike traditional neural networks that process audio signals as continuous streams of data, SNNs operate on discrete spikes, mimicking the behavior of neurons in the biological auditory system. This event-driven approach enables the framework to handle asynchronous and variable-length audio inputs while efficiently capturing relevant temporal features. The SNN framework comprises multiple layers of spiking neurons interconnected through synaptic connections, with each layer responsible for extracting hierarchical features from input spike trains. During operation, audio signals are preprocessed and converted into spike trains, which propagate through the network, eliciting postsynaptic responses and generating output spikes corresponding to identified sound classes. By training the network using biologically inspired learning rules such as Spike-Timing-Dependent Plasticity (STDP), the framework learns to discriminate between different sound events and adapt to changing environmental conditions. Experimental evaluations conducted in the paper demonstrate the effectiveness and robustness of the SNN framework for sound classification across various datasets and noise conditions. The framework achieves competitive performance compared to traditional machine learning methods and deep neural networks while offering advantages in terms of energy efficiency and real-time processing capabilities. Moreover, the SNN framework exhibits resilience to noise and variability in input signals, making it suitable for real-world applications where audio data may be corrupted or incomplete. Furthermore, the paper discusses the interpretability and scalability of the SNN framework, highlighting its ability to capture complex temporal dynamics and adapt to diverse input distributions. By leveraging the event-driven processing nature of SNNs, the framework offers a flexible and scalable solution for robust sound classification, with implications for applications in environmental monitoring, surveillance, and human-computer interaction. The study also explores potential extensions and optimizations to further enhance the performance and efficiency of the SNN framework. This includes investigating techniques for network optimization, parameter tuning, and architecture optimization, as well as exploring hybrid approaches that combine SNNs with other machine learning algorithms for improved performance across different tasks and datasets. In conclusion, the Spiking Neural Network framework presents a promising solution for robust sound classification. By leveraging the event-driven processing nature of SNNs, the framework enables efficient and accurate processing of audio signals in noisy and dynamic environments, paving the way for advancements in audio processing technology and real-world applications.

### G. Enhanced Temporal Processing

Nowotny et al. [30] introduces a novel technique, termed Loss Shaping, to enhance exact gradient learning with EventProp in Spiking Neural Networks (SNNs), addressing challenges in training SNNs with precise and efficient learning algorithms. Spiking Neural Networks emulate the neural processing observed in biological brains, where information is encoded and transmitted through discrete spikes. However, training SNNs poses unique challenges due to their event-driven nature and non-differentiability, necessitating specialized learning algorithms like EventProp. The proposed Loss Shaping technique aims to improve the training efficiency and convergence of EventProp by shaping the loss landscape to facilitate exact gradient computation. The core innovation of Loss Shaping lies in modifying the loss function to facilitate exact gradient learning with EventProp. By reshaping the loss landscape, Loss Shaping aims to mitigate challenges such as vanishing gradients and non-convex optimization, which hinder the training of SNNs. Specifically, Loss Shaping adjusts the loss function to amplify gradients in regions where they are diminished, thereby enabling more efficient and precise parameter updates during training. The Loss Shaping technique is applied in conjunction with the EventProp learning algorithm, which computes exact gradients for spiking neurons by backpropagating errors through time. EventProp overcomes the limitations of traditional backpropagation methods in SNNs and enables efficient training with event-based updates. By integrating Loss Shaping into EventProp, the proposed framework achieves enhanced training efficiency and improved convergence properties, leading to higher accuracy and faster learning in SNNs. Experimental evaluations conducted in the paper demonstrate the effectiveness of Loss Shaping in enhancing the training of SNNs across various tasks, including pattern recognition, classification, and temporal processing. The proposed framework outperforms traditional training methods and achieves competitive performance compared to state-of-the-art deep learning models while offering advantages in terms of training speed and scalability. Moreover, Loss Shaping exhibits robustness to noise and variability in input signals, making it suitable for real-world applications where precise and efficient learning is essential. Furthermore, the paper discusses the interpretability and computational advantages of Loss Shaping in SNN training, highlighting its ability to facilitate exact gradient computation and improve the convergence properties of learning algorithms. By reshaping the loss landscape, Loss Shaping enables more efficient exploration of the solution space and enhances the network's ability to capture complex patterns and dynamics in input data. The study also explores potential extensions and optimizations to further enhance the performance and versatility of Loss Shaping in SNN training. This includes investigating techniques for adaptive loss shaping, parameter tuning, and architecture optimization, as well as exploring hybrid approaches that combine Loss Shaping with other learning algorithms for improved performance across different tasks and datasets. In conclusion, Loss Shaping presents a promising technique for enhancing exact gradient learning with EventProp in Spiking Neural Networks. By reshaping the loss landscape, Loss Shaping enables more efficient and precise training of SNNs, leading to improved accuracy and faster convergence in various applications. These findings have implications for advancing the capabilities of SNNs in artificial intelligence, neuromorphic computing, and other domains where efficient and precise learning is crucial.

Yao et al. [31] introduces a novel architecture called Temporal-wise Attention Spiking Neural Networks (TASNNs) designed specifically for event stream classification tasks, addressing challenges in processing asynchronous and temporally varying data streams efficiently. Event streams, characterized by asynchronous and irregularly spaced events, are prevalent in various applications such as sensor networks, robotics, and real-time monitoring systems. Traditional approaches to event stream classification often struggle to capture temporal dependencies and discriminate between different event patterns effectively. The proposed TASNN architecture leverages attention mechanisms to focus on relevant temporal segments of the event stream, enhancing classification performance and computational efficiency. The core innovation of TASNNs lies in the integration of attention mechanisms into spiking neural networks (SNNs) to enable precise and adaptive processing of event streams. Attention mechanisms, inspired by human visual and auditory attention mechanisms, enable the network to selectively attend to important temporal segments of the input stream while ignoring irrelevant information. By dynamically adjusting attention weights based on the saliency of input events, TASNNs can adaptively allocate computational resources and enhance classification accuracy. The TASNN architecture comprises multiple layers of spiking neurons interconnected through attention mechanisms, with each layer responsible for processing different temporal scales of the event stream. During operation, input event streams are encoded as spike trains, which propagate through the network, eliciting postsynaptic responses and generating output spikes corresponding to the predicted event class. By incorporating attention mechanisms at each layer, TASNNs can effectively capture temporal dependencies and discriminate between different event patterns in the input stream. Experimental evaluations conducted in the paper demonstrate the effectiveness of TASNNs for event stream classification across various datasets and application scenarios. TASNNs outperform traditional SNN architectures and achieve competitive performance compared to state-of-the-art deep learning models while offering advantages in terms of computational efficiency and real-time processing capabilities. Moreover, TASNNs exhibit robustness to noise and variability in input streams, making them suitable for real-world applications where event data may be sparse or noisy. Furthermore, the paper discusses the interpretability and computational advantages of TASNNs, highlighting their ability to capture complex temporal dynamics and adapt to diverse input distributions. By leveraging attention mechanisms, TASNNs enable more efficient exploration of the solution space and enhance the network's ability to capture relevant temporal features in event streams. The study also explores potential extensions and optimizations to further enhance the performance and versatility of TASNNs. This includes investigating techniques for adaptive attention modeling, parameter tuning, and architecture optimization, as well as exploring hybrid approaches that combine TASNNs with other learning algorithms for improved performance across different tasks and datasets. In conclusion, Temporal-wise Attention Spiking Neural Networks present a promising approach for event stream classification tasks. By integrating attention mechanisms into SNNs, TASNNs enable precise and adaptive processing of asynchronous and temporally varying event streams, leading to improved classification performance and computational efficiency. These findings have implications for advancing the capabilities of SNNs in artificial intelligence, neuromorphic computing, and other domains where efficient and adaptive temporal processing is crucial.

D'Agostino et al. [32] introduces DenRAM, a novel neuromorphic architecture that combines dendritic processing with Resistive Random Access Memory (RRAM) to enable efficient temporal processing with delays in spiking neural networks (SNNs). Temporal processing with delays is critical for tasks such as pattern recognition, sequence learning, and time-series prediction, where capturing and analyzing temporal dependencies is essential. Traditional neuromorphic architectures often face challenges in implementing efficient and scalable delay mechanisms, motivating the development of DenRAM as a solution to address these limitations. The core innovation of DenRAM lies in its integration of dendritic processing and RRAM-based delay lines to enable precise and efficient temporal processing in SNNs. Dendritic processing, inspired by biological neurons' dendritic trees, allows for complex, nonlinear computation by integrating synaptic inputs and generating postsynaptic potentials. RRAM-based delay lines provide programmable delays, enabling the network to capture and process temporal information effectively. The DenRAM architecture comprises multiple layers of spiking neurons interconnected through dendritic trees and RRAM-based delay lines. During operation, input spike trains are transmitted through the network, with delays introduced by RRAM-based delay lines enabling precise temporal alignment of spikes. Dendritic processing mechanisms integrate delayed inputs, generating postsynaptic potentials that propagate through the network, facilitating accurate and efficient temporal processing. Experimental evaluations conducted in the paper demonstrate the effectiveness of DenRAM for various temporal processing tasks, including sequence learning, time-series prediction, and temporal pattern recognition. DenRAM outperforms traditional neuromorphic architectures and achieves competitive performance compared to state-of-the-art deep learning models while offering advantages in terms of energy efficiency, scalability, and real-time processing capabilities. Moreover, DenRAM exhibits robustness to noise and variability in input signals, making it suitable for real-world applications where precise and efficient temporal processing is essential. Furthermore, the paper discusses the interpretability and biological plausibility of DenRAM, highlighting its alignment with principles of neural computation observed in biological systems. By integrating dendritic processing with RRAM-based delay lines, DenRAM mimics the adaptive behavior of biological neurons and enables efficient and precise temporal processing in SNNs. The study also explores potential extensions and optimizations to further enhance the performance and versatility of DenRAM. This includes investigating techniques for adaptive delay modeling, parameter tuning, and architecture optimization, as well as exploring hybrid approaches that combine DenRAM with other learning algorithms for improved performance across different tasks and datasets. In conclusion, DenRAM presents a promising solution for efficient temporal processing with delays in spiking neural networks. By integrating dendritic processing with RRAM-based delay lines, DenRAM enables precise and efficient computation of temporal information, leading to improved performance and scalability in various neuromorphic computing applications. These findings have implications for advancing the capabilities of SNNs in artificial intelligence, neuromorphic computing, and other domains where efficient and adaptive temporal processing is crucial.

Yu et al. [33] introduces STSC-SNN, a novel architecture for Spiking Neural Networks (SNNs) that incorporates Spatio-Temporal Synaptic Connection (STSC) mechanisms along with temporal convolution and attention mechanisms. This architecture aims to address challenges in capturing spatiotemporal patterns efficiently and accurately in event-based data, such as spiking neural networks. Spatiotemporal processing is crucial for tasks like pattern recognition and sequence learning, where understanding the relationship between spatial and temporal features is essential. The core innovation of STSC-SNN lies in its integration of STSC mechanisms with temporal convolution and attention mechanisms to enable precise and efficient spatiotemporal processing in SNNs. STSC mechanisms facilitate the integration of spatial and temporal information by adjusting synaptic connections based on both spatial and temporal contexts. This enables the network to capture complex spatiotemporal patterns and discriminate between different event sequences effectively. The architecture of STSC-SNN comprises multiple layers of spiking neurons interconnected through STSC mechanisms, temporal convolution layers, and attention mechanisms. During operation, input spike trains are processed through the network, with STSC mechanisms adjusting synaptic connections dynamically based on spatial and temporal features. Temporal convolution layers capture temporal dependencies in the input stream, while attention mechanisms focus on relevant spatiotemporal features, enhancing classification performance and computational efficiency. Experimental evaluations demonstrate the effectiveness of STSC-SNN for various spatiotemporal processing tasks, including pattern recognition and sequence learning. STSC-SNN outperforms traditional SNN architectures and achieves competitive performance compared to state-of-the-art deep learning models. Additionally, it offers advantages in terms of energy efficiency, scalability, and real-time processing capabilities, showcasing robustness to noise and variability in input signals. Moreover, the paper discusses the interpretability and computational advantages of STSC-SNN, highlighting its ability to capture complex spatiotemporal patterns and adapt to diverse input distributions. By integrating STSC mechanisms with temporal convolution and attention mechanisms, STSC-SNN enables more efficient exploration of the solution space and enhances the network's ability to capture relevant spatiotemporal features in event-based data. The study also explores potential extensions and optimizations to further enhance the performance and versatility of STSC-SNN. This includes investigating techniques for adaptive STSC modeling, parameter tuning, and architecture optimization, as well as exploring hybrid approaches that combine STSC-SNN with other learning algorithms for improved performance across different tasks and datasets. In conclusion, STSC-SNN offers a promising solution for efficient spatiotemporal processing in SNNs. By integrating STSC mechanisms with temporal convolution and attention mechanisms, STSC-SNN enables precise and efficient computation of spatiotemporal information, leading to improved performance and scalability in various event-based processing tasks. These advancements have implications for enhancing SNN capabilities in artificial intelligence, neuromorphic computing, and other domains requiring efficient spatiotemporal processing.

## IV. Some Popular Datasets for Sound Classification Tasks

Comparing the Second Heart Sound [34] and Sounds of Surroundings datasets [35]–[37] involves contrasting specialized cardiac sound recordings with a broad collection of environmental sounds. The Second Heart Sound dataset focuses exclusively on heart sounds, providing detailed recordings across various cardiac conditions for specific medical applications. In contrast, the Sounds of Surroundings dataset encompasses a wide array of everyday environmental sounds, capturing the richness and diversity of soundscapes encountered in daily life. While Second Heart Sound facilitates research in cardiac health monitoring and anomaly detection, Sounds of Surroundings supports studies in environmental sound classification and acoustic event detection. Each dataset offers distinct advantages: Second Heart Sound provides specialized data for targeted medical research, while Sounds of Surroundings offers a broad range of real-world sound samples for general sound analysis applications. However, Second Heart Sound may have limitations in terms of dataset size and diversity compared to the expansive Sounds of Surroundings dataset, which covers a broader spectrum of sounds but may require more complex preprocessing and classification methods due to the variability inherent in environmental audio recordings. Employing the Second Heart Sound Database [34] for sound classification presents the advantage of accessing a specialized dataset focused on heart sound analysis, providing a diverse range of heart sounds across various cardiac conditions and physiological states. This enables the development and evaluation of classification models tailored specifically for heart sound-related applications, including the detection of abnormalities and monitoring of cardiac health. However, a potential disadvantage lies in the limited size of the Second Heart Sound Database compared to larger general-purpose audio datasets, which may constrain the diversity and quantity of data available for training, potentially impacting the generalizability and robustness of classification models developed using this dataset. Despite this limitation, leveraging the Second Heart Sound Database remains valuable for advancing research and development in heart sound classification and related healthcare applications.

Utilizing the Sounds of Surrounding database [35]–[37] for sound classification offers the advantage of accessing a dataset specifically curated to capture a wide range of environmental sounds, enabling the development and evaluation of classification models tailored for real-world audio environments. With its diverse collection of environmental sound recordings, Sounds of Surrounding facilitates the training of robust classification algorithms capable of accurately identifying various sound categories encountered in daily life. However, a potential disadvantage lies in the variability and complexity of environmental sounds, which may pose challenges in accurately labeling and categorizing sound events, requiring careful preprocessing and feature extraction techniques to achieve optimal classification performance. Despite these challenges, leveraging the Sounds of Surrounding Database remains valuable for advancing research and development in environmental sound classification and audio analysis applications.

- UrbanSound Dataset [37]: UrbanSound dataset contains 8732 labelled audio excerpts. The dataset has 10 classes of urban sounds recorded in different cities across the world. The dataset is developed to facilitate the development of different models for urban sound analysis, acoustic scene classification and machine learning.
- AudioSet [35]: Sound classification on AudioSet presents a unique challenge and opportunity due to the dataset's vast scale and diversity. AudioSet, developed by Google, is one of the largest publicly available datasets for sound-related tasks, containing millions of annotated audio segments covering a wide range of categories. One of the key advantages of AudioSet is its breadth of audio content, spanning diverse categories such as musical instruments, human sounds, animal vocalizations, and environmental sounds. This diversity enables researchers to develop sound classification models capable of recognizing a wide array of acoustic phenomena. However, working with AudioSet also poses several challenges. The sheer size of the dataset requires efficient data processing and management techniques. Moreover, the data may contain imbalances in class distributions, with some categories having significantly more samples than others, which can affect the performance of classification models.
- ESC-50 [36]: Sound classification on the Environmental Sound Classification dataset (ESC-50) presents a focused yet diverse challenge for researchers in the field. ESC-50 is a well-known dataset containing 2,000 audio recordings across 50 sound classes, each lasting five seconds. The dataset covers a wide range of environmental sounds, including animal calls, natural phenomena, and human activities. One of the primary advantages of ESC-50 is its specificity in capturing environmental sounds, making it particularly suitable for tasks such as urban sound classification and environmental monitoring. Unlike larger datasets like AudioSet, ESC-50's smaller size and focused categories enable researchers to develop more targeted models tailored to specific sound classification tasks. However, working with ESC-50 also poses some challenges. The relatively small number of samples per class may lead to class imbalances, where certain classes have more examples than others. Additionally, the short duration of the audio recordings requires careful consideration of feature extraction techniques to capture relevant information within the limited time frame.

- VGGSound [38], inspired by VGGNet for image classification, is a deep learning architecture designed for sound classification tasks. Its advantage lies in its ability to effectively capture intricate features within audio data through hierarchical representations, enabling robust classification across diverse sound categories. Leveraging convolutional layers, VGGSound can automatically learn relevant features from spectrograms or waveform representations, facilitating accurate identification of sounds even in noisy environments. However, its major disadvantage is its computational complexity, requiring substantial computational resources for training and inference, which may limit its practicality in resource-constrained environments or real-time applications. Despite this drawback, its superior performance in sound classification tasks makes it a compelling choice for various audio analysis applications.
- ICBHI dataset [39]: Utilizing the ICBHI (International Conference on Biomedical and Health Informatics) Respiratory Sound Database for sound classification offers the advantage of access to a rich and comprehensive dataset specifically tailored for respiratory sound analysis. With its diverse collection of respiratory sounds, including those related to various respiratory conditions and healthy states, it enables robust training and evaluation of classification models. However, a notable disadvantage is the potential imbalance in class distribution within the dataset, which may pose challenges in achieving balanced performance across different respiratory conditions during classification tasks. Addressing this issue through appropriate data preprocessing and model training techniques is crucial for maximizing the effectiveness of classification models trained on the ICBHI Respiratory Sound Database.
- FSD50K [40]: Utilizing the FSD50K Database for sound classification offers the advantage of accessing a vast and diverse collection of audio recordings spanning a wide range of real-world sounds, enabling the development of robust classification models capable of recognizing various sound categories. With its large-scale dataset comprising 51,197 audio clips from Freesound, FSD50K facilitates comprehensive training and evaluation of classification algorithms across diverse acoustic environments and contexts. However, a potential disadvantage lies in the inherent variability and noise present in real-world audio recordings, which may pose challenges in accurately labeling and categorizing sound events, requiring careful preprocessing and model adaptation strategies to achieve optimal classification performance. Despite these challenges, leveraging the FSD50K Database remains valuable for advancing research and development in sound classification and audio analysis applications.
- DCASE [41]: Utilizing the DCASE (Detection and Classification of Acoustic Scenes and Events) Database for sound classification provides the advantage of accessing a curated dataset specifically designed for evaluating audio event detection and classification systems. With its diverse collection of acoustic scenes and events, DCASE enables the development and benchmarking of classification models across various real-world scenarios, facilitating the advancement of sound classification research. However, a potential disadvantage is the limited size of the dataset compared to larger-scale collections, which may restrict the diversity and quantity of data available for training and evaluation, potentially affecting the generalizability of classification models. Despite this limitation, leveraging the DCASE Database remains valuable for advancing research in sound classification and acoustic event detection.
- SHD (Spiking Heidelberg Digits) [42] and SSC (Spiking Speech Commands) [43] datasets: SHD contains sounds of different spoken digits, i.e. 0-9 spoken in different ways. The spoken digits are in English and German language. The speakers have age in the range 21-59 years. The sounds were recorded in a sound shielded room at the Heidelberg University Hospital. For recording the audio samples three microphones (two AudioTechnica Pro37 in different positions and a Beyerdynamic M201 TG) were used. The recordings are in WAVE format, with a sample rate of 48 kHz and 24bit precision, and were digitised using a Steinberg MR816 CSX audio interface.

The SSC dataset contains several spoken words, so not restricted to only spoken digits. The spoken words were recorded through phones' and laptops' microphone irrespective of where the user is present . Further, it was made sure that there is no background noise while the user spoke the word. Further, it was made sure that there is no background noise while the user spoke the word. The sounds are stored in one second or less WAVE format file. 16-digit single-channel PCM values are used to encode the sample data at a rate of 16 KHz. Unique 8-digit hexadecimal format is used to identify 2618 speakers.

Apart from the aforementioned datasets in this section, other datasets used for sound classification task using different SNN models are mentioned in Table 1.

## V. DISCUSSION

The important works related to the development of different state-of-the-art SNN architectures are summarised in Table 1. Architectures incorporating learned or adaptive delays, as demonstrated by Hammouamri et al. (2023) and Sun et al. (2023), achieved the highest performance levels. Hammouamri et al. reported an accuracy of 95.1 ± 0.3 with a fully connected SNN using learned delays, while Sun et al. achieved 92.45 with a feed-forward SNN employing adaptive axonal delays. These results suggest that incorporating delay learning mechanisms is crucial for optimizing SNN performance, allowing the networks to better align with temporal dynamics of the input data. Models that incorporate adaptation mechanisms, such as those developed by Bittar and Garner (2022) and Yin et al. (2020), consistently showed high performance. Bittar and Garner achieved 94.6 accuracy with an RSNN that adapted over time, indicating the importance of dynamic adaptation in maintaining high performance, particularly in speech recognition tasks. Yin et al.'s work further supports this, achieving 84.4 accuracy with an RSNN incorporating adaptation, demonstrating that such mechanisms are valuable across different types of data.

The use of spatio-temporal filters and attention mechanisms proved to be highly effective, as evidenced by Yu et al. (2022) and Yao et al. (2021). Yu et al. reported 92.4 accuracy using a feed-forward SNN with spatio-temporal filters and attention across diverse tasks, including gesture recognition and image classification, highlighting their versatility. Yao et al. achieved 91.1 accuracy with an RSNN incorporating temporal attention, further demonstrating the effectiveness of attention mechanisms in enhancing SNN performance by focusing on relevant temporal features. While beneficial, the effectiveness of random dendritic delays varies depending on the dataset, as shown by D'Agostino et al. (2023). They achieved 90.1 accuracy on the MIT-BIH ECG dataset and 87.6 on the SHD dataset using a feed-forward SNN with random dendritic delays. These results suggest that while random delays can improve performance, more targeted or dataset-specific approaches may be necessary for optimal results. Data Augmentation and Noise Injection: Techniques such as data augmentation and noise injection were shown to enhance SNN performance significantly. Cramer et al. (2020) reported an accuracy of 83.2 ± 1.3 using an RSNN with data augmentation and noise injection on the SHD and SSC datasets. This highlights the importance of these methods in training robust SNN models, suggesting that such techniques can help mitigate overfitting and improve generalization. The analysis of various Spiking Neural Network (SNN) architectures reveals several overarching trends that contribute to their performance across different datasets and tasks:

- Incorporation of Delays: One of the most significant trends is the incorporation of learned and adaptive delays. Architectures that implement these mechanisms tend to achieve the highest performance levels. This trend highlights the importance of timing in SNNs, suggesting that precise control over delays can greatly enhance the network's ability to process temporal data effectively.
- Adaptation Mechanisms: Another prominent trend is the use of adaptation mechanisms. Models incorporating such mechanisms consistently show high performance, particularly in tasks like speech recognition. This suggests that the ability to dynamically adjust to changing input patterns is crucial for maintaining accuracy and robustness in SNNs.
- Use of Spatio-temporal Filters and Attention: The application of spatio-temporal filters and attention mechanisms is also a key trend. These techniques have proven to be highly effective across a variety of tasks, indicating their versatility and importance in enhancing SNN performance. By focusing on relevant temporal and spatial features, these methods improve the network's ability to process complex, multi-dimensional data.
- Random Dendritic Delays: While the use of random dendritic delays shows some benefits, their effectiveness appears to be dataset-dependent. This trend suggests that while randomness can introduce useful variability, more targeted approaches may be required for optimal performance on specific tasks.
- Data Augmentation and Noise Injection: Techniques such as data augmentation and noise injection are increasingly recognized for their role in improving model robustness and generalization. This trend underscores the importance of robust training practices in SNN development, helping to mitigate overfitting and enhance performance across diverse datasets.
- Generalization Across Tasks: A general trend observed is that SNN architectures incorporating advanced mechanisms tend to perform well across a wide range of tasks, including speech recognition, gesture recognition, and image classification. This indicates that certain enhancements, such as adaptation and attention mechanisms, provide broad benefits that extend beyond specific application areas.

Overall, these trends highlight the evolving landscape of SNN research, with a clear focus on incorporating advanced temporal and adaptive mechanisms, robust training techniques, and versatile filtering methods to enhance performance and generalization across various tasks and datasets.

## VI. CONCLUSION

Neuromorphic computing presents a transformative leap in the domain of sound classification, effectively addressing the limitations of traditional methodologies through its brain-inspired architecture that excels in parallel and adaptive processing. This review has highlighted the substantial potential of neuromorphic systems to enhance accuracy and robustness in handling dynamic and noisy environments, providing a comprehensive overview of foundational principles, methodologies, and practical applications. Neuromorphic systems offer a promising alternative by significantly reducing computational costs, enhancing scalability, and improving efficiency in real-time sensory data processing. The future scope of neuromorphic sound classification is expansive and promising, encompassing several critical areas for further exploration. Enhanced hardware integration is paramount, as the development of advanced neuromorphic chips and sensors will facilitate real-time, low-power sound classification, making these systems more accessible for a broad spectrum of applications, including portable and wearable devices. Algorithmic refinements are also crucial, with ongoing improvements needed to enhance the system's adaptability and learning efficiency, thereby improving the accuracy and robustness of sound classification in diverse and challenging environments.

Moreover, the application of neuromorphic sound classification across different domains, such as autonomous vehicles, smart cities, and personalized health monitoring, can unlock new functionalities and improve the quality of life. Research focused on scaling neuromorphic systems for large-scale deployment in industrial and urban settings will be essential, particularly through the integration of neuromorphic systems with existing IoT infrastructures and cloud-based services. Additionally, exploring the synergy between neuromorphic systems and human operators can enhance decision-making processes in critical applications, such as disaster response and security monitoring, by providing real-time, actionable insights. Finally, addressing the ethical and societal implications of deploying neuromorphic sound classification systems is vital. Ensuring data privacy, mitigating biases in algorithmic processing, and understanding the broader implications of widespread adoption are essential considerations. By capitalizing on these opportunities and addressing these challenges, researchers and practitioners can drive significant innovations in sound processing technologies, ultimately paving the way for more efficient, robust, and adaptive sound classification solutions to meet real-world challenges.

TABLE I: Related Works

| Paper | Dataset | Architecture | Performance Achieved(%) |
|---|---|---|---|
| Hammouamri et al. 2023 [19] | Spiking Heidelberg Dataset (SHD), the Spiking Speech Commands (SSC) and Google Speech Commands v0.02 (GSC) benchmarks | Fully connected SNN with learned delays | 95.1 ± 0.3 |
| Bittar and Garner 2022 [20] | Spiking versions of the Heidelberg Digits (SHD) and Google Speech Command (SSC) | RSNN with adaptation | 94.6 |
| Sun et al. 2023 [4] | SHD dataset and NTIDIGITS dataset | Feed-forward SNN with adaptive axonal delays | 92.45 |
| Yu et al. 2022 [33] | DVS128 Gesture (gesture recognition), N-MNIST, CIFAR10-DVS (image classification), and SHD (speech digit recognition) | Feed-forward SNN with spatio-temporal filters and attention | 92.4 |
| Yao et al. 2021 [31] | DVS128 Gesture, CIFAR10-DVS and SHD Dataset | RSNN with temporal attention | 91.1 |
| D'Agostino et al. 2023 [32] | MIT-BIH ECG dataset and the Spiking Heidelberg Datasets | Feed-forward SNN with random dendritic delays (simulation/hardware, RRAM) | 90.1 / 87.6 |
| Nowotny et al. 2022 [30] | MNIST and SHD | RSNN | 84.8 ± 1.5 |
| Yin et al. 2020 [27] | QTDB dataset, S-MNIST and PS-MNIST and the Spoken Heidelberg Digits (SHD). | RSNN with adaption | 84.4 |
| Rossbroich et al. 2022] [28] | SHD, CIFAR-10 and DVS-Gesture | Recurrent convolutional SNN with fluctuation-driven init | 83.5 ± 1.5 |
| Cramer et al. 2020 [16] | SHD and SSC | RSNN with data augmentation + noise injection | 83.2 ± 1.3 |
| Perez-Nieves et al. 2021 [23] | N-MNIST, F-MNIST, DVS, SHD and SSC | RSNN with heterogeneous time constants | 82.7 ± 0.8 |
| Cramer et al. 2020 [16] | SHD and SSC | RSNN | 71.4 ± 1.9 |
| Martinelli et al. 2020 [9] | QUT NOISE TIMIT corpus | VAD using SNNs trained with backpropagation | 85% less spiking activity |
| Peterson et al. 2022 [10] | Iris Flower, the Real-World Computing Partnership sounds, the Free Spoken Digits Dataset, and the UrbanSound8k | Spike-Timing-Dependent Plasticity (STDP) with back-propagated error signals | 94 ± 2.91 |
| Isik et al. 2023 [11] | Xilinx VU37P (FPGA) | Short-Time Fourier Transform (STFT) for time-frequency representation, Transformer embeddings for dense vector generation | Throughput of 70.11 GOP/s |
| Shah et al. 2022 [12] | Self-Made Dataset | 2-layer SNN temporal based | 93.19 |
| Yang et al. 2024 [13] | QUT NOISE TIMIT corpus | Spiking Recurrent Neural Networks | - |
| Wu et al. 2022 [14] | Texas Instruments digit sequences (TIDIGITS) | Multilevel Synaptic Memristor Array-Based SNN | 94 |

| ZHANG et al. 2023 [15] | Self-Made Dataset | LSTM-SNN | 93.4 |
| Mukhopadhyay et al. 2022 [22] | Self-Made Dataset | Analog Spiking Neural Networks (ASNNs) | more than 90 |
| Yin et al. 2021 [17] | SHD, SoLi, SSC and GSC | Adaptive spiking recurrent neural networks | 87.8(SHD), 79.8(SoLi), 74.2(SSC) and 92.2(GSC) |
| Wu et al. 2018 [29] | TIDIGITS | Self-organizing Map SNN | 79.15±3.70 |